\documentclass[12pt,epsf]{article}
\usepackage{epsfig}
\usepackage{amssymb}  
\usepackage{amsthm}
\usepackage{amsmath}
\let\eqref\cref

\renewcommand{\thanks}[1]{\footnote{#1}} 

\newcommand{\be}{\begin{equation}}
\newcommand{\ee}{\end{equation}}
\newcommand{\bea}{\begin{eqnarray}}
\newcommand{\eea}{\end{eqnarray}}
\newcommand{\half}{{1\over 2}}

\def\stacksymbols#1#2#3#4{\def\theguybelow{#2}
\def\verticalposition{\lower#3pt}
\def\spacingwithinsymbol{\baselineskip0pt\lineskip#4pt}
\mathrel{\mathpalette\intermediary#1}}
\def\intermediary#1#2{\verticalposition\vbox{\spacingwithinsymbol
\everycr={}\tabskip0pt
\halign{$\mathsurround0pt#1\hfil##\hfil$\crcr#2\crcr
\theguybelow\crcr}}}

\newcommand{\goesto}[2]{\stacksymbols{\Longrightarrow}{{}_{#1 \rightarrow #2}}{4}{0.1} }

\begin{document}

\pagestyle{empty}

\bigskip\bigskip
\begin{center}
{\bf \large A Plausible Path Towards Unification of Interactions via Gauge Fields Consistent with the Equivalence Principle-I}
\end{center}

\begin{center}
James Lindesay\footnote{E-mail address, jlindesay@howard.edu} \\
Computational Physics Laboratory \\
Howard University,
Washington, D.C. 20059 
\end{center}
\bigskip

\begin{abstract} 

An extension of the Lorentz group to include
generators $\Gamma^\mu$ carrying a space-time index is demonstrated to 
\emph{explicitly} construct the Minkowski metric \emph{within} the internal group space as a 
consequence of the non-vanishing commutation relations between those generators. 
For fundamental representation states, this group forms a subgroup of GL(4) consistent with Dirac fermions. 
Since the representations inherently pair particles with anti-particles, the generators
define a minimal group for transforming fields that fundamentally
satisfy microscopic causality, thus treating causality as a crucial physical property.

For fermions transforming under GL(4),
beyond three spin matrices and $\Gamma^0$ there are 12 additional 
hermitian generators.
It is demonstrated that a closed invariance subgroup of SU(2)$\times$U(1) for fermions of defined mass
can be constructed.  However, any closed subgroup of SU(3) transformations \emph{necessarily} mix
the mass-SU(2) eigenstates. The first part of exploration of group representations presented in this paper
examines associating \emph{global} invariance to these additional generators. 
The extended group requires that any charges arising due to establishment of \emph{internal} local symmetries necessarily
exhibit geometric equivalence under curvilinear transformations on the parameterization of space-time translations.
Furthermore, a formulation for mixing the degenerate
massless fermions (which explicitly \emph{cannot} manifest SU(3) symmetry or charges)
while \emph{maintaining} Lorentz invariance, is presented.

An exploration of the first boson representation of this algebra will be presented in the second paper. 
Of particular interest, the degenerate components of that representation can be mixed
consistent with the standard electro-weak model in a manner that will be shown to provide potential
insights into kinematic relationships between electro-weak bosons.  Such suggestive relationships could
demonstrate a path towards enhancing understandings of constraints on the properties
of such emergent states.

\end{abstract}

\bigskip 

\setcounter{equation}{0}
\section{Introduction}
\indent \indent

The formulation of relativistic dynamics by Dirac\cite{Dirac} introduced charge-conjugate
spinor components whose dynamical equations were linear in the energy-momentum operators
(versus the quadratic form of the Klein-Gordon equation). 
The resulting resolvents (propagators) then invert linear operators,
considerably simplifying the analytic structure of formulations for causal scattering behaviors and
cluster decomposability of the fermions. Cluster decomposability is of course crucial for
quantum-to-classical correspondence. 
Furthermore, the inherent pairing of conjugate particles enables the fundamental representations of
Dirac \emph{fields} to directly satisfy
requisite properties of microscopic causality\cite{WeinbergQFT}. 
A generalized formulation that requires the form  $\hat{\Gamma}^\mu \: \hat{P}_\mu$  to
be a Lorentz scalar operation, while satisfying the Dirac-like equation
\be
\mathbf{\Gamma}^\mu \cdot {\hbar \over i} { \partial \over \partial x^\mu} \,
\hat{\mathbf{\Psi}}_{\gamma}^{(\Gamma)}
(\vec{x}) = -\gamma m  c  \: \hat{\mathbf{\Psi}}_{\gamma}^{(\Gamma)}(\vec{x}),
\label{CausalSpinorFieldEqn}
\ee
has been developed\cite{JLFQG} to result as the dynamical equation
within a closed group of physical transformations.  
For the $\hat{\mathbf{\Psi}}_{\gamma}^{(\Gamma)}$ fields
(which will be referred to as \emph{causal fields}), the masses are
always non-negative, and for massive particles the extra factor $\gamma$ is the eigenvalue of the hermitian
operator  $\hat{\Gamma}^0$ which is \emph{not} positive semi-definite. 
For most formulations of unitary causal scattering theory, the single mass/energy parameter
associated with eigenvalues of the fundamental equation of dynamics goes ``off-shell"
during the ``off-diagonal" coherent superpositions of a complete set of intermediate states\cite{JLScattering,AlexThesis},
and overall energy-momentum conservations results from ``on-shell" final state limits in this parameter
(along with the various $e^{{i \over \hbar} \vec{p} \cdot \vec{x}}$ factors).
Under Poincaré transformations, the field satisfies
\be
\hat{U}(\mathbf{\Lambda},\vec{a})\left [ \hat{\mathbf{\Psi}}_{\gamma}^{(\Gamma)}(\vec{x}) \right ]_b \hat{U}^\dagger (\mathbf{\Lambda},\vec{a}) =
\sum_{b'} \mathcal{D}_{b b'}(\mathbf{\Lambda}^{-1}) \left [ \hat{\mathbf{\Psi}}_{\gamma}^{(\Gamma)}(\mathbf{\Lambda}\vec{x}+\vec{a}) \right ]_{b'},
\label{PoincareTransformEqn}
\ee
where $\mathcal{D}_{b b'}(\mathbf{\Lambda})$ are finite dimensional representations of the general Lorentz group
transformation $\mathbf{\Lambda}$. 
The Dirac equation is proportional to the
fundamental  $\Gamma=\half$ finite dimensional (fermionic) representations for the fields in (\ref{CausalSpinorFieldEqn}), with
the standard  4$\times$4 Dirac matrices  $\gamma^\mu$ satisfying $\mathbf{\Gamma}^\mu=\half \gamma^\mu$.

The inclusion of non-abelian generators $\hat{\Gamma}^\mu$ that 
directly carry a space-time index $\mu$ provides (through the \emph{non-vanishing} structure constants) 
a group-theoretical mechanism for
directly generating the Minkowski metric $\eta_{\alpha\beta}$ as a group metric between generators (as occurs for Casimir operators),
which could not be done using the abelian generators of infinitesimal space-time translations $\hat{P}_\mu$ alone. 
Such invariants,
along with interactions associated with internal symmetry groups of (\ref{CausalSpinorFieldEqn}),
become incorporated in standard gravitation via the principle of equivalence and 
curvilinear coordinate transformations on the abelian (flat) space-time translations. 
The physical manifestation of the three fundamental \emph{dimensional} constants $\hbar, \, c$, and $G$ distinguish
gravitation as distinct from the gauge couplings between interacting systems. 
General relativity continues to successfully describe the macro-physical phenomena of gravitation,
as well as the gravitation of coherent quantum states\cite{PDG2022, Overhauser,MullerPetersChu},
by coupling the purely geometric space-time curvature to the local energy-momentum density. 
Non-gravitational measurables are integrated through equivalence on locally flat (freely falling)  coordinates $\xi^\alpha(x)$
with a Minkowski metric form
$\eta_{\alpha \beta}= {\partial x^\mu \over \partial \xi^\alpha}
g_{\mu \nu}(x) {\partial x^\nu \over \partial \xi^\beta}$. 

The generators of the Lorentz group $\hat{J}_j$ and $\hat{K}_j$, along with the generators $\hat{\Gamma}^\mu$,
form a closed subgroup of the group of linear transformations GL(4). 
The sixteen hermitian generators of GL(4) include the three angular momentum spin matrices,
$\Gamma^0$,  and 12 additional matrices that have the same number of degrees
of freedom as SU(3)$\times$SU(2)$\times$U(1).  Any unitary subgroups within the GL(4) that can be associated
with interactions and charges inherently share relationships between their Lie structure functions\cite{Hamermesh}
(which are functions of the group parameters) that ultimately define the structure constants. 
If an invariance (sub)group of
parameters take on a local \emph{space-time} dependency, the mapping  of the parameter dependencies onto compact
space-time surfaces define quantization conditions on any charges associated with
that (sub)group\cite{SBthesis,JLHLM,JLFQG}.  Thus, one expects that relationships
between any shared charged carriers of local subgroups isomorphic to internal symmetries within
the GL(4) would be implemented into gravitating environments via the
principle of equivalence, which was the original motivation for this approach.

In what follows,
fundamental group properties relevant to the discussion will be summarized. 
Subsequently, closed subgroups of the additional hermitian generators of GL(4) that leave stationary
standard state spinor fields for fundamental ($\Gamma=\half$) fermions invariant will be explored. 
Any local gauge fields that one can associate with `internal' symmetries will have a 
\emph{single} space-time index as well as
completely independent matrix indexes that transform under those group transformations,
thereby establishing equivalence associations on curvilinear extensions of the space-time translations. 
Interactions between particles manifest as bosonic fields or environs (e.g. electro-weak bosons, gluons,
gravitation,...).  The subsequent paper (part II) will explore
$\Gamma=1$ spinors as potential descriptors of electro-weak bosons.


\setcounter{equation}{0}
\section{Properties of a Closed Group Inclusive of $\mathbf{\Gamma}^\mu$} 
\indent 

The closed algebra that extends the Lorentz group via the
inclusion of the operators $\hat{\Gamma}^\mu$ satisfies the following commutation relations:
\be
\begin{array}{lll}
\left [ \Gamma^0 \, , \, \Gamma^k \right] \: = \: {i \over \hbar} \, K_k  , &
\left [ \Gamma^0 \, , \, J_k \right] \: = \: 0  , &
\left [ \Gamma^0 \, , \, K_k \right] \: = \: -i  \hbar \,  \Gamma^k  , \\
\left [ \Gamma^j \, , \, \Gamma^k \right] \: = \: -{ i \over \hbar} \, \epsilon_{j k m} \, J_m  , &
\left [ \Gamma^j \, , \, J_k \right] \: = \: i \hbar \, \epsilon_{j k m} \, \Gamma^m  , &
\left [ \Gamma^j \, , \, K_k \right] \: = \: -i \hbar \, \delta_{j k} \, \Gamma^0  .
\end{array}
\label{ExtLorentzGroupEqns}
\ee
The operators $J_k$ and $\Gamma^0$ are hermitian, while the operators $K_k$ and $\Gamma^k$
are anti-hermitian. 
The Casimir invariant that labels the irreducible representations of this group is given by
$C_\Gamma \: = \: {1 \over 6} \left [ \,  ( \underline{J} \cdot \underline{J} \,-\, \underline{K} \cdot \underline{K} )/\hbar^2
\,+\, \Gamma^0 \, \Gamma^0 \,-\, \underline{\Gamma} \cdot \underline{\Gamma} \right ]$.
Representation labels will be attributed to the
eigenvalues of the mutually commuting operators $C_\Gamma$,  $\Gamma^0$, $J^2$, and  $J_3$,
given by ${2 \Gamma (\Gamma + 2)  \over 6}, \: \gamma, \: J(J+1) \hbar^2$, and $J_z \hbar$, respectively.
Finite dimensional representations have dimension
$N_\Gamma \: = \: {1 \over 3} (\Gamma + 1) (2 \Gamma + 1) (2 \Gamma + 3)$.
The labels satisfy $0 \le J \le \Gamma$,
$-J \le \gamma \le J$, and $-J \le J_z \le J$, thus
the signatures (half-integral vs. integral) of $\Gamma$, $J$, $\gamma$, and $J_z$ are necessarily the same.
In general, the operators in (\ref{ExtLorentzGroupEqns}) can have system+internal components
$\hat{\Xi}_r=\hat{\chi}_r+\hat{X}_r$, (like orbital and spin angular momentum),
where $[\hat{\chi}_r,\hat{X}_m]=0$, with $\mathbf{X}_m$ denoting matrix representations of the internal group
operators.  For angular momentum, the symbols $(s,s_z)$ will be used for the \emph{internal} spin
angular momenta.

Fundamental matrices corresponding to $\Gamma={1 \over 2}$ 
 have dimensionality $N_{1 \over 2}=4$,
with $\mathbf{\Gamma^0}=\half \gamma^0_{Dirac}$ and $\mathbf{J}_3$ both diagonal.
Thus, the algebra (\ref{ExtLorentzGroupEqns}) forms a subgroup of GL(4).
The general spinor forms of the operators satisfying the commutation relations (\ref{ExtLorentzGroupEqns})
can be found in \cite{JLFQG}.

\subsection{Development of a group metric on space-time indexes}

For a general group algebra
$\left [ \hat{G}_r \, , \, \hat{G}_s \right ] \: = \: -i \, \sum_m \left ( c_s \right ) _r ^m \, \hat{G}_m $
with appropriately non-vanishing structure constants $ c_{s r}{}^m=- c_{r s}{}^m$,
those constants can be used to construct a group metric $\eta_{a b}$ on the generators of the algebra\cite{Hamermesh},
\be
\eta_{a b} \: \equiv \: \sum_{s \, r} \left ( c_a \right )_r ^s \, \left ( c_b \right )_s ^r, \:
\eta^{ab}\equiv  ((\eta)^{-1})_{ab} .
\ee
This metric defines invariants using products of group generators, such as the 
Casimir operator $\hat{C}_\mathcal{G} \equiv \sum_{r s} \eta^{G_r G_s} \hat{G}_r \hat{G}_s$.
The non-commuting operators $\hat{\Gamma}^\mu$ that carry a space-time index
paired with the 4-momentum operator in (\ref{CausalSpinorFieldEqn}) define a group metric of Lorentz sub-group invariants given by
\be
\eta^{\Gamma^\mu \, \Gamma^\nu} \: = \: -{1 \over 6} \, \eta_{\mu \, \nu},
\ee
where $\eta_{\mu \, \nu}$ is the usual Minkowski metric. This metric later generates invariants involving
the energy-momentum generators (as tangent vectors on a curvilinear manifold) when the group is extended to include space-time translations.
Thus, the Minkowski metric is \emph{explicitly} generated within this closed algebra,
beyond its \emph{implicit} involvement in describing properties of Lorentz transformations. 

Any physical observables that satisfy a group algebra must be consistent with the
group structure.  One therefore expects that
measurable hermitian observables within a \emph{physically} fundamental
GL(4) share a group connection with the actions of the algebra on
space-time translations, and thus (via the principle of equivalence) measurements in
curvilinear space-times.  Such observables (e.g. spin)
must be consistent with the general group structure that \emph{includes}
those translations.  In terms of the \emph{abelian} space-time translations,
any general Lie group sequential 4-translations given by $\vec{a}$ then $\vec{b}$ within a
Poincaré extension of (\ref{ExtLorentzGroupEqns}) with group (closure) operation
$\vec{\Phi}(\vec{b},\underline{b};\vec{a},\underline{a})$
can be re-expressed in terms of `locally-flat' coordinates satisfying\cite{JLFQGch9}
\be
\xi^\mu (\vec{\Phi}(\vec{b},\underline{I};\vec{a},\underline{I}))=\xi^\mu(\vec{b})+\xi^\mu(\vec{a}),
\ee
where the $\underline{I}$ represents the identity transformation in the other group parameters. The tetrads
that connect the curvilinear metric to the Minkowski metric (mentioned in the Introduction) satisfy
${\partial \xi^\mu(\vec{x})\over \partial x^\beta}=
\left . {\partial \Phi^\mu(\vec{x}^{-1},\underline{G};\vec{x}',\underline{I}) \over \partial x'^{\beta}} \right |_{\vec{x}'\rightarrow \vec{x}}^{ \underline{G}\rightarrow \underline{I}}$.

\subsection{Form of spinor fields}

General causal spinor fields can be chosen to satisfy normalization expressed in the form
\bea
\hat{\mathbf{\Psi}}_{(\gamma)}^{(\Gamma)}(\vec{x})={1 \over \sqrt{2}} \sum_{s_z}\int d^3 p \,
{m\,c^2 \over \epsilon(\underline{p})}
\left [ {e^{{i \over \hbar}(\underline{p} \cdot \underline{x}-\epsilon(\underline{p})t)} \over (2 \pi \hbar)^{3/2}} 
 \mathbf{u}_{\gamma}^{(\Gamma)}(\underline{\beta},s,s_z) 
\hat{a}_{\gamma}^{(\Gamma) }(\vec{p},s,s_z) \: + \right . \quad \quad \nonumber \\
\left.  (-)^{s+s_z} {e^{{-i \over \hbar}(\underline{p} \cdot \underline{x}-\epsilon(\underline{p})t)} \over (2 \pi \hbar)^{3/2}}
 \mathbf{u}_{-\gamma}^{(\Gamma)}(\underline{\beta},s,-s_z) 
\hat{a}_{-\gamma}^{(\Gamma) \dagger}(\vec{p},s,s_z)
\right ] \equiv \sum_{s_z}\int d^3 p \,
{m\,c^2 \over \epsilon(\underline{p})} \hat{\mathbf{\Phi}}_{(\gamma)\underline{p}}^{(\Gamma)}(\vec{x}),
\label{CausalFieldEqn}
\eea
where for massive particles the momenta can be expressed in terms of the (active) Lorentz transformation from
the standard energy state $\vec{p}=(\gamma_L m_{(st)} c,\gamma_L \underline{\beta}\: m_{(st)} c)$,
and the Lorentz factor given by $\gamma_L=\sqrt{1 \over 1-|\underline{\beta}|^2}$ should not be confused
with the eigenvalue of $\mathbf{\Gamma}^0$.

The standard state, which \emph{always} corresponds to $|\underline{\beta}| \rightarrow 0$, has (contravariant) 4-momentum
components $\vec{p}_{(st)}=(m_{(st)} c,0,0,0)$ for massive particles,
and traditionally takes the z-moving (helicity) form
$(1,0,0,1){\epsilon_{(st)}\over c}$ for massless particles.  The spinor
$\mathbf{u}_{\gamma}^{(\Gamma)}(\mathbf{\underline{\beta}},s,s_z)$ is dimensionless and
appropriately normalized.  Any preferred normalization factors (e.g ${m c^2 \over \epsilon(\underline{p})} d^3 p
\goesto{\underline{p}}{0}d^3 p$) can be absorbed in redefined creation operators. 
The action of $\mathbf{\Gamma}^\mu p_\mu$ on a spinor yields
\be
\mathbf{\Gamma}^\mu p_\mu \mathbf{u}_{\gamma}^{(\Gamma)}(\underline{\beta},s,s_z) = 
-\gamma \, m_{(st)} c \, \mathbf{u}_{\gamma}^{(\Gamma)}(\underline{\beta},s,s_z),
\label{SpinorEquation}
\ee
where $\underline{\beta}={\underline{p}\over \epsilon(|\underline{p}|)}$.

A hermitian spinor metric $\mathbf{g}_\Gamma$ connecting the hermitian adjoint to the Dirac adjoint field
$\bar{\mathbf{\Psi}} \equiv \mathbf{\Psi}^\dagger \mathbf{g}_\Gamma$
satisfying (\ref{CausalSpinorFieldEqn}) can be developed.  This matrix has
the same dimension $N_\Gamma$ as the matrices
$\mathbf{X}_s  \in \{ \mathbf{J}_j, \mathbf{K}_k, \mathbf{\Gamma}^\mu \}$, and must satisfy
${\mathbf{X}_s}^\dagger = \mathbf{g}_\Gamma \mathbf{X}_s \mathbf{g}_\Gamma$.
Equation (\ref{CausalSpinorFieldEqn}) then implies
${\partial \over \partial x^\mu}\left [ \hat{\bar{\mathbf{\Psi}} }_{(\gamma)}^{(\Gamma)}(x)
\mathbf{\Gamma}^\mu \hat{\mathbf{\Psi}}_{(\gamma)}^{(\Gamma)}(x) \right ]=0$
as  a local conservation condition. 
The matrix $\mathbf{g}_\Gamma$ is purely diagonal, with elements $(-1)^{\Gamma-\gamma}$. 
For the fundamental representation, the spinor metric is also related to one of the Dirac matrices
$\mathbf{g}_\half=\gamma^0$.  However, this is no longer the case for representations $\Gamma >\half$.


\setcounter{equation}{0}\
\section{Properties of Fundamental Representation Fermions}
\indent

In this section, the  $\Gamma=\half$ representation fermions will be demonstrated. 
As previously mentioned, the Dirac matrices satisfy $\gamma^\mu=2\Gamma^\mu$.
The Dirac-normalized 4-spinors have upper (`particle') and lower (`anti-particle') components
given by
 \be
\mathbf{u}_{\half}^{(\half)}(\underline{\beta},\half,s_z)=\sqrt{\gamma_L+1 \over 2}
\left ( \begin{array}{c} \chi(s_z) \\ {\underline{\sigma} \cdot \underline{\beta} \over \gamma_L +1} \chi(s_z)  \end{array} \right ), \:
\mathbf{u}_{-\half}^{(\half)}(\underline{\beta},\half,s_z)=\sqrt{\gamma_L+1 \over 2}
\left (  \begin{array}{c} {\underline{\sigma} \cdot \underline{\beta} \over \gamma_L +1} \chi(s_z) \\  \chi(s_z)   \end{array} \right ),
\label{4SpinorGeneralForm}
\ee
where the 2-spinors are defined by $\chi(\half)=\left ( \begin{array}{c} 1 \\ 0 \end{array} \right )$ and
$\chi(-\half)=\left ( \begin{array}{c} 0 \\ 1 \end{array} \right )$,
and $\mathbf{\underline{\sigma}}$ are the Pauli spin matrices.  The normalization satisfies
$\mathbf{u}_{\gamma}^{(\half)}(\underline{\beta},\half,s_z)^\dagger \gamma^0
\mathbf{u}_{\gamma'}^{(\half)}(\underline{\beta},\half,s_z)=\textnormal{sign}(\gamma) \delta_{\gamma \gamma'}$.
For standard Dirac 4-spinors\cite{Kaku}, $\mathbf{u}(\vec{p},s_z)=\mathbf{u}_{\half}^{(\half)}(\underline{\beta},\half,s_z)$ and
$\mathbf{v}(\vec{p},s_z)=\mathbf{u}_{-\half}^{(\half)}(\underline{\beta},\half,-s_z)$.

\subsection{Additional Hermitian Generators in GL(4) \label{HermitianGenerators}}

Next, complete sets of all hermitian generators in GL(4) will be developed.  Along with the 3 angular
momenta $\mathbf{S}_j$ and $\mathbf{\Gamma^0}$, there are 12 additional hermitian matrices (including
the identity).  Since this number corresponds to the number of generators in SU(3)$\times$SU(2)$\times $U(1),
one is motivated to explore combinations of hermitian matrices that leave a given standard state invariant,
since these matrices generate global symmetries of that state\cite{JLSU3SU2U1}.

For the present, the focus will be on symmetries of a ($s_z=+\half,\gamma=+\half$)
\emph{massive} particle spinor $\mathbf{u}_{+(st)}(+\half)$,
which (in the rest frame) is a column spinor with a single non-vanishing component of 1 as the first component. 
For the general spinor form in (\ref{4SpinorGeneralForm}),
it is straightforward to develop a transformation matrix $\mathbf{L}(\underline{\beta})$
from a standard-state massive spinor through
$\mathbf{u}_{\gamma}^{(\half)}(\underline{\beta},\half,s_z)\equiv
\mathbf{L}(\underline{\beta})\mathbf{u}_{\gamma(st)}(s_z)$. 
If the particle with standard state $\mathbf{u}_{+(st)}(+\half)$ moves parallel to the z-axis, this matrix takes the form
\be
\mathbf{L}(\beta_z)=
{1 \over \sqrt{2}}\left( \begin{array}{cccc}  \sqrt{1+\gamma_L}&0&{\gamma_L \beta_z \over \sqrt{1+\gamma_L}}&0 \\
0&1&0&0 \\ {\gamma_L \beta_z \over \sqrt{1+\gamma_L}} & 0 & \sqrt{1+\gamma_L} & 0 \\
0&0&0&1
\end{array}\right).
\ee
The Dirac normalization relationship expressible in the form $\mathbf{L}^{-1}=\gamma^0 \mathbf{L}^\dagger \gamma^0$
insures that $\mathbf{L}(\underline{\beta})$ is invertible. 
By examination of the standard state (and even z-moving spinors) one can develop a closed
SU(2) invariance sub-group on the spinors of the form
\be
\tau_1\equiv \half \left(
\begin{array}{cccc} 0&0&0&0 \\ 0&0&0&1 \\ 0&0&0&0 \\ 0&1&0&0
\end{array}
\right),\:
\tau_2\equiv \half \left(
\begin{array}{cccc} 0&0&0&0 \\ 0&0&0&-i \\ 0&0&0&0 \\ 0&i&0&0
\end{array}
\right),\:
\tau_3\equiv \half \left(
\begin{array}{cccc} 0&0&0&0 \\ 0&1&0&0 \\ 0&0&0&0 \\ 0&0&0&-1
\end{array}
\right).
\ee
The massive \emph{helicity} state spinors are \emph{generally} invariant under transformations using these generators,
independent of momentum, as are also the massless state spinors.

Next, a closed representation of SU(3) that leaves $\mathbf{u}_{+(st)}(+\half)$ invariant will be developed.
The 3-dimensional representation of SU(3) has 8 generators that will be denoted by $\mathbf{Q}_r$. 
These generators will be embedded in the 4$\times$4 space
using $\mathbf{T}_r\equiv \left( \begin{array}{cc} 0 &\underline{\mathbf{0}}^T\\
\underline{\mathbf{0}}&\mathbf{Q}_r \end{array}\right )$, where $\underline{\mathbf{0}}$
denotes a 3-D null column matrix.  Clearly these generators satisfy the SU(3) algebra, as well as
the invariance relation $\mathbf{T}_r \mathbf{u}_{+(st)}(+\half)=\mathbf{0}$. 
In a commonly utilized representation of SU(3), the generators
$\mathbf{Q}_3$ and $\mathbf{Q}_8$ are diagonal, so that along with $\mathbf{1}$,
$\mathbf{\Gamma}^0$, $\mathbf{S}_z$, and
$\tau_3$, these 6 generators \emph{over} count the number of diagonal hermitian matrices in GL(4), which means they
cannot be linearly independent.  To develop sets of linearly independent matrices in the desired form,
the following extra matrices will be defined:
\bea
\mathbf{M}_3\equiv \half \left(
\begin{array}{cccc} 0&0&1&0 \\ 0&0&0&0 \\ 1&0&0&0 \\0&0&0&0
\end{array}
\right),\:
\mathbf{M}_4\equiv \half \left(
\begin{array}{cccc} 0&0&-i&0 \\ 0&0&0&0 \\i&0&0&0 \\ 0&0&0&0
\end{array}
\right), \nonumber\\
\mathbf{M}_5\equiv \half \left(
\begin{array}{cccc} 0&0&0&-i \\ 0&0&0&0 \\ 0&0&0&0 \\i&0&0&0
\end{array}
\right),\:
\mathbf{M}_8\equiv \half \left(
\begin{array}{cccc} 0&0&0&1 \\ 0&0&0&0 \\0&0&0&0 \\ 1&0&0&0
\end{array}
\right).
\eea

Utilizing various sets of the previously defined matrices, one can
develop a complete set of 16 hermitian generators inclusive of the SU(2) invariance subgroup
given by
$\{\mathbf{1},\mathbf{S}_x,\mathbf{S}_y,\mathbf{S}_z,\mathbf{\Gamma}^0,\tau_1,\tau_2,\tau_3,
\mathbf{T}_1,\mathbf{T}_2,\mathbf{M}_3,\mathbf{M}_4,\mathbf{M}_5,\mathbf{T}_6,\mathbf{T}_7,\mathbf{M}_8  \}$. 
The remaining SU(3) generators satisfy
$\mathbf{T}_3=\half \left( \mathbf{\Gamma}^0 - \mathbf{S}_z \right )
-\tau_3, \mathbf{T}_4=\tau_1, \mathbf{T}_5=-\tau_2$, and $
\mathbf{T}_8=-\half\left( \mathbf{\Gamma}^0 - \mathbf{S}_z \right ) -\tau_3$. 
Alternatively, one can develop generators inclusive of the SU(3) invariance subgroup given by
$\{\mathbf{1},\mathbf{S}_x,\mathbf{S}_y,\mathbf{S}_z,\mathbf{M}_3,\mathbf{M}_4,\mathbf{M}_5,\mathbf{M}_8$,
$\mathbf{T}_1,\mathbf{T}_2,\mathbf{T}_3,\mathbf{T}_4,\mathbf{T}_5,\mathbf{T}_6,\mathbf{T}_7,\mathbf{T}_8  \}$
which \emph{necessarily} mix $\mathbf{\Gamma}^0$-mass-SU(2) eigenstates.
The absent generators decompose as follows: $\mathbf{\Gamma}^0=\mathbf{S}_z +\mathbf{T}_3 -\mathbf{T}_8,
\tau_1=\mathbf{T}_4,\tau_2=-\mathbf{T}_5$, and $\tau_3=-\half\left( \mathbf{T}_3 +\mathbf{T}_8 \right)$. 
For completeness,
expressed in terms of the standard set of Dirac matrices $\gamma^\mu, \:\gamma^5\equiv i \gamma^0\gamma^1
\gamma^2\gamma^3,\:\gamma^5 \gamma^\mu$ and $\sigma^{\mu \nu}\equiv {i \over 2}[\gamma^\mu,\gamma^\nu]$,
the additional generators decompose as follows: $\sigma^{j 0}=2 \mathbf{K}_j, \: \sigma^{j k}=2 \epsilon_{j k m}\mathbf{S}_m, \:
\gamma^5=2 \tau_1+2 \mathbf{M}_3=2 \mathbf{T}_4 +2 \mathbf{M}_3,
\: \gamma^5 \gamma^1=-2 \mathbf{S}_x +4 \mathbf{T_1},\:
\gamma^5 \gamma^2=-2 \mathbf{S}_y-4\mathbf{T}_2$, and $\gamma^5 \gamma^3=-2 \mathbf{\Gamma}^0+4 \tau_3=
-2 \mathbf{S}_z-4 \mathbf{T}_3$.

For \emph{either} of the special unitary sub-groups, the general \emph{boosted} spinor
(\ref{4SpinorGeneralForm}) remains invariant under the
independent subgroups $\mathbf{G}_r (\underline{\beta})\equiv
\mathbf{L}(\underline{\beta})\mathbf{G}_r \mathbf{L}(\underline{\beta})^{-1}$. 
The similarity transformation maintains the commutation relations, and if it is done on the whole set of
hermitian generators (or even just those exclusive of angular momentum),
the linear independence is preserved.  However, the transformation generally takes the
SU(3) generators $\mathbf{T}_s$ out of the 3-D invariance subspace.

\subsection{Constraints on Hermitian Symmetries of Fundamental Fermions}

Next, consider a massive fermion field of the form
(\ref{CausalFieldEqn}), which satisfies a \emph{local} gauge symmetry on
hermitian GL(4) generators $\mathbf{G}_b$.  Generally,
the introduction of gauge fields allows interactions to be introduced via minimal coupling,
where the various $e^{{i \over \hbar} \vec{p} \cdot \vec{x}}$ factors guarantee appropriate energy-momentum conservation after de-coherences of co-moving entanglements.

\subsubsection{Massive fermion states}

Examine the equation of motion
\be
\mathbf{\Gamma}^\mu \cdot \left [ \mathbf{I}  {\hbar \over i} { \partial \over \partial x^\mu} 
-{q_b \over c} A_\mu^b(\vec{x}) \mathbf{G}_b \right ]\,
\hat{\mathbf{\Psi}}_{(\gamma)}^{(\Gamma)}(\vec{x})
= -\gamma m  c  \: \hat{\mathbf{\Psi}}_{(\gamma)}^{(\Gamma)}(\vec{x}),
\label{GaugeInvariantSpinorFieldEqn}
\ee
where here and henceforth repeated indices are assumed to be summed over.
Under the \emph{local} gauge transformation
$\tilde{\mathbf{\Psi}}(\vec{x})=\mathbf{U(\alpha (\vec{x}))}\mathbf{\Psi}(\vec{x})$
that also transforms the spinors, one expects the operation on the field equation to be homogeneous, i.e.
\be
\mathbf{U}(\alpha(\vec{x})) \mathbf{\Gamma}^\mu \left [ \mathbf{I}  {\hbar \over i} { \partial \over \partial x^\mu} 
-{q_b \over c} \tilde{A}_\mu^b(\vec{x}) \mathbf{G}_b \right ] =
 \mathbf{\Gamma}^\mu  \left [ \mathbf{I}  {\hbar \over i} { \partial \over \partial x^\mu} 
-{q_b \over c} A_\mu^b(\vec{x}) \mathbf{G}_b \right ] \mathbf{U}(\alpha(\vec{x})),
\ee
which defines a relationship between the two expressions of the gauge field
\be
-{q_b \over c} \tilde{A}_\mu^b(\vec{x})  \mathbf{\Gamma}^\mu \mathbf{G}_b =
\mathbf{U}^{-1}(\alpha(\vec{x}))  {\hbar \over i}  \mathbf{\Gamma}^\mu { \partial \over \partial x^\mu} \mathbf{S}(\alpha(\vec{x}))
-{q_b \over c} A_\mu^b(\vec{x}) \mathbf{U}^{-1}(\alpha(\vec{x})) \mathbf{\Gamma}^\mu \mathbf{G}_b  \mathbf{S}(\alpha(\vec{x})).
\ee
Under infinitesimal transformations where 
$\mathbf{S}(\delta \alpha^b(\vec{x})) \simeq \mathbf{1}+i\delta \alpha^b(\vec{x}) \mathbf{G}_b$, this gives
\be
{q_b \over c} \tilde{A}_\mu^b(\vec{x}) \mathbf{\Gamma}^\mu \mathbf{G}_b \simeq
{q_b \over c} A_\mu^b(\vec{x}) \mathbf{\Gamma}^\mu \mathbf{G}_b  
+{q_b \over c} A_\mu^b(\vec{x})  \delta \alpha^a (\vec{x}) [\mathbf{\Gamma}^\mu \mathbf{G}_b,\mathbf{G}_a]
-\hbar  { \partial \over \partial x^\mu} \delta \alpha^b (\vec{x})  \mathbf{\Gamma}^\mu \mathbf{G}_b.
\label{InfinitesimalGT}
\ee
Thus, the gamma matrices would be involved in such GL(4) transformations.

The massive plane wave components have spinors that satisfy
\be
\mathbf{\Gamma}^\mu \left[  p_\mu 
- {q_b \over c} A_\mu^b(\vec{x}) \mathbf{G}_b \right ] \mathbf{u}_{\gamma}^{(\Gamma)}(\underline{\beta},s,s_z)  = 
-\gamma \, m c \, \mathbf{u}_{\gamma}^{(\Gamma)}(\underline{\beta},s,s_z) ,
\ee
where for boundary states, $ A_\mu^b(\vec{x})  \rightarrow 0$.  Thus one is motivated to examine whether
the special unitary subgroups of the GL(4) can establish charges defining product spaces related to
the observed SU(3)$\times$SU(2)$\times$U(1) internal symmetry spaces, whose generators are independent
of the extended Lorentz transformations.

From the previous spinor example (\ref{4SpinorGeneralForm})
one should note that $e^{i\chi^j \mathbf{\mathbf{\tau}}_j }\mathbf{u}_{\pm(st)}(\pm\half)=
\mathbf{u}_{\pm(st)}(\pm\half)$ (with correlated signs)
and $e^{i\xi^r \mathbf{\mathbf{T}}_r }\mathbf{u}_{+(st)}(+\half)=\mathbf{u}_{+(st)}(+\half)$
are \emph{invariant} GL(4) transformations on the standard state examined.  Such invariances can be
developed on any of the 4 massive standard state spinors. 
Furthermore, for \emph{arbitrary} z-moving spinors (i.e. helicity states), the action of
$[p_\mu \mathbf{\Gamma}^\mu \tau_j,\tau_k]$ on $\mathbf{u}_{+}(\beta_z,+\half)$
vanishes,
thus maintaining a general global invariance under that representation of generators of SU(2). 
This remains true also for specified
helicity \emph{massless} states, as will be discussed later.  One can generally demonstrate that either
$[\mathbf{\Gamma}^\mu \tau_j,\tau_k]=0$ or
$[\mathbf{\Gamma}^\mu \tau_j,\tau_k]\propto\mathbf{\Gamma}^\beta \tau_m$.
More generally for
$\mathbf{G}_b \in \{\tau_j,\mathbf{T}_r \}$, the transformation
$[p_\mu \mathbf{\Gamma}^\mu \mathbf{G}_a(\underline{\beta}),\mathbf{G}_b(\underline{\beta})]
\mathbf{u}_{+}(\underline{\beta},+\half)=0 $ for \emph{arbitrary} Lorentz
boosts from the standard state, and
$\mathbf{u}_{\pm}(\underline{\beta},\pm\half)\rightarrow
e^{i \chi^b (\vec{x}) \tau_b (\underline{\beta})} \:
\mathbf{u}_{\pm}(\underline{\beta},\pm\half)$ (with correlated signs) leaves
(\ref{GaugeInvariantSpinorFieldEqn}) invariant for arbitrary local transformations $\chi^b (\vec{x})$. 
There is \emph{always} a specific Lorentz frame associated 
with a specified initial state $\underline{\beta_o}$. 
This then defines the symmetry group associated with coherent intermediate propagation.  However, the two sets $\{\mathbf{\Gamma}^0,\mathbf{\tau}_j\}$ and 
$\{ \mathbf{T}_r \}$ cannot \emph{simultaneously} establish linearly independent behaviors.

The spinor of any \emph{initial} state
maintains this invariance during coherent mixing with intermediate states, while the 
off-diagonal energy-momentum and space-time relationships (due to intermediate state completeness)
remain uncertain during off-shell propagation and interactions. 
Furthermore, it can be shown that the vacuum expectation value $\left< 
\hat{\bar{\mathbf{\Phi}}}_{(\gamma)\underline{p'}}^{(\Gamma)}(\vec{x}) \tau_b
\hat{\mathbf{\Phi}}_{(\gamma)\underline{p}}^{(\Gamma)}(\vec{x}) 
\right >_{vacuum}=0$ vanishes for any of the SU(2) generators for all momenta.  
If this symmetry expresses as an internal local symmetry on
the fields $\hat{\mathbf{\Psi}}_{(\gamma)}^{(\Gamma)}(\vec{x})$ and develops \emph{charges}
associated with that symmetry (forming an isomorphic
product space associated with those charges), this allows the exchange of quanta with other
systems expressing such charges.

Due to confinement
of bound SU(3) symmetric fields, one expects the corresponding bound state fields analogous to (\ref{CausalFieldEqn})
to manifest \emph{specific} particle or antiparticle behaviors within the bound states, suggesting that the
aforementioned SU(3) exemplars would behave similarly (by not requiring mixing of particle and antiparticle
components within those states).  Relativistic formulations of bound particle states based upon
extensions of the Dirac's equation (with additional complications) have been developed
by others\cite{Dirac1950,CraterEtAl,Sazdjian}.
Distinct particle or antiparticle components of bound states are expected to have a spinor component that
allows a proper CKM-SU(3) symmetry analogous to that described in section \ref{HermitianGenerators}.
Thus, the formulation here presented is suggestive that the observed
CKM-SU(3)$\times$SU(2)$\times$U(1) local symmetry groups describing observed phenomena
is a consequence of the set of GL(4) invariances that can act upon the standard state spinors of the extended
Lorentz group representations.

A primary motivation of this approach comes from the recognition that the circulations or fluxes of gauge
fields through compact space-time surfaces (closed curves, surfaces, etc.) should be completely calculable
from the group composition functions $\phi^s(\underline{\beta};\underline{\alpha})$ defining
subgroup closure within the full group of GL(4) transformations.  For instance, the flow fields of both isotropic and
anisotropic textured superfluids define gauge potentials with quantized charges that quantify
the interactions between flow fields when the mapping of the group topology onto space-time
surfaces is no longer simply connected (e.g. between vortex states, terminating textures,
etc.)\cite{JLHMGQF,SBthesis,JLHLM}. 
In an analogous manner, one should be able
to examine topological mappings onto compact space-time surfaces of the gauge potentials
associated with the previously discussed symmetries to determine charge quantization relationships
(see e.g. sections 3.4 and 4.2 in reference \cite{JLFQG}).

\subsubsection{Massless fermion states}

The finite dimensional internal representation of $\mathbf{\Gamma}^\mu$ will \emph{always} commute
with the momentum operator $\hat{P}_\nu$, while still transforming under Lorentz transformations.  However,
if the operators $\hat{\Gamma}^\mu$ have both internal (finite dimensional) and system-wide actions (like spin and
orbital angular momentum), then the Jacobi identity involving $\hat{P}_\nu$, $\hat{K}_j$, and $\hat{\Gamma}^\mu$
necessitates at least one additional operator $\hat{\mathcal{M}}_A$ (for group closure) that satisfies
\be
[\hat{\Gamma}^\mu,\hat{P}_\nu]=i \delta_\nu^\mu \hat{\mathcal{M}}_A c , \quad
[\hat{\Gamma}^\mu,\hat{\mathcal{M}}_A c]=i\eta^{\mu \nu} \hat{P}_\nu ,
\label{AffineMassCommutes}
\ee
with the extended Poincare group invariant given by $\hat{\mathcal{P}}^2=\hat{\mathcal{M}}_A^2 c^2-\eta^{\mu \nu}
\hat{P}_\mu \hat{P}_\nu$.  The standard state energies $\epsilon_{(st)}$ (e.g. $m c^2$) serve as invariants labeling
the particle states that transform under representations of this extended Poincare group.

Using (\ref{AffineMassCommutes}), one can show that\cite{JLFQG}
\bea
e^{{i\over \hbar} \vec{a}\cdot \hat{\vec{P}}} \hat{\Gamma}^\mu \hat{P}_\mu e^{-{i\over \hbar} \vec{a}\cdot \hat{\vec{P}}} =
\hat{\Gamma}^\mu \hat{P}_\mu +{a^\mu \hat{P}_\mu \over \hbar} \hat{\mathcal{M}}_A c, \nonumber \\
e^{{i\over \hbar} \xi \hat{\mathcal{M}}_A c} \hat{\Gamma}^\mu \hat{P}_\mu e^{-{i\over \hbar} \xi \hat{\mathcal{M}}_A c} =
\hat{\Gamma}^\mu \hat{P}_\mu +{\xi \hat{P}_\mu \eta^{\mu \nu} \hat{P}_\nu \over \hbar}.
\eea
Invariance of the $\hat{\Gamma}^\mu \hat{P}_\mu$ operation under space-time translations
requires that  the affine mass eigenvalue $\mathcal{M}_A$ vanishes for massive states, and that 
massive particles cannot manifest finite affine
translations $\xi\neq 0$ generated by $\hat{\mathcal{M}}_A$.  The inner product of 
the translation of a light-like system or wave packet $(c \Delta t,\Delta \underline{x})$ with its 4-momentum
$p^{\mu}\rightarrow ({\epsilon \over c},\underline{p})$ also vanishes.
Thus if there is an external component to $\hat{\Gamma}^\mu$,
just as rest mass defines the scale associated with infinitesimal translations along the particle's proper time,
analogously the affine mass generates affine translations $\xi$ along a massless particle's light-cone trajectory. 
Inertial masses are inherently energy-like, while systems with non-vanishing affine mass are inherently
light-like.  Affine translations along light cones are likely related to light-cone parameters found
in the literature\cite{Dirac1948,LSJL2004,BrodskyPauli}.

The established formulations of electromagnetism require that \emph{all} photons have the same
affine mass (i.e. standard state energy $\epsilon_{(st)}\equiv \mathcal{M}_\gamma c^2$, typically
assigned a value of 1 energy unit for standard electromagnetic coupling), defining a
single type of coupling with interacting particles. 
However, this extension of the Poincare group
provides a mechanism for mixing non-identical \emph{massless} particle states that do not violate
Lorentz invariance, since energies and momenta remain conserved during detections. 
Relativistic formulations that utilize Lorentz invariant phase space descriptions would incorporate
factors ${\epsilon_{(st)} \over \epsilon(\underline{p})}d^3 p$
in completeness integrals over intermediate states analogous to those utilized for massive particles\cite{RedShelf} .

Massless standard states are traditionally defined as helicity states moving parallel to the z-axis,
and the standard state spinors take the form
\be
u_{(\pm \half)}^{(\half)}(\vec{p}_{(st)},\half,+\half)={1 \over \sqrt{2}}\left(
\begin{array}{c} \pm 1\\0\\ 1\\0 \end{array} \right ), \quad
u_{(\pm \half)}^{(\half)}(\vec{p}_{(st)},\half,-\half)={1 \over \sqrt{2}}\left(
\begin{array}{c} 0\\ \pm 1\\0\\ -1 \end{array} \right ),
\ee
independent of the momentum.  As mentioned before, the SU(2) subgroup of GL(4) previously developed
continues to be an invariance subgroup for the $s_z =+\half$ standard states, and a commensurate set
can be developed for the $s_z =-\half$ standard states. 
However, \emph{no} linearly independent invariance SU(3) subgroup can be developed for the \emph{massless}
fundamental representation fermions. Thus, helicity $\pm \half$ massless fermions are not expected
to be able to carry definitive charges generated
from an internal SU(3) invariance within the GL(4), which is consistent with neutrino phenomenology.

\subsubsection{Mixing of massless fermions}

The equation of motion (\ref{CausalSpinorFieldEqn}) is degenerate for \emph{all} massless causal fermions.  This allows
degenerate massless fermions to mix during coherent propagation for which relationships between
affine translations and energy invariants remain uncertain.  
 Lorentz invariance is insured as long as the (non-dispersive) mixing probabilities are independent of the Lorentz frame,
even if standard state energy labels differ.

Coherently propagating eigenstates will be labeled using the group invariant affine mass $\left |\mathcal{M}_{r} \right >$, and
interaction-detected states will be labeled $\left | d_a \right >$. 
Since detection breaks coherence via an interaction,
mixed initial and final state eigenstates are expected to differ by a propagation phase according to
\be
\left| d_a \right > = \sum_r \left |\mathcal{M}_{r} \right > U_{r a} , \quad
\left< d_b | d_a \right > = \sum_r U_{r b}^* e^{-{i \over \hbar} \mathcal{M}_{r} c \Delta \xi} U_{r a}.
\ee
Defining $\Upsilon_{s r}(b,a)\equiv U_{s b}U_{s a}^*U_{r b}^* U_{r a}$
(with both real and imaginary parts, analogous to usual formulations
of neutrino mixing)\cite{MNSnumixing,BPnumixing,NakamuraPetcov}
and $\Delta \mathcal{M}_{s r} \equiv \mathcal{M}_s -\mathcal{M}_r$, the mixing probabilities
$\mathcal{P}_{ba}\equiv \left | \left< d_b | d_a \right > \right |^2$ take the form
\be
\mathcal{P}_{ba}=\delta_{b a} - 2 \sum_{s>r} \left \{
2 \Re (\Upsilon_{s r}(b,a)) \sin^2 \left ({\Delta \mathcal{M}_{s r}c \Delta\xi \over 2 \hbar}\right ) -
\Im  (\Upsilon_{s r}(b,a)) \sin \left ({\Delta \mathcal{M}_{s r}c \Delta\xi \over \hbar}\right )
\right \}.
\label{NeutrinoMixingEqn}
\ee

If there is coherent propagation through differing types of interacting media, there can be ``dispersive" differences in how
an eigenstate of a particular generation propagates through a particular media type.  For instance, 
the mixing of atmospheric neutrinos is seen to differ significantly during propagation within sparse versus dense baryonic media
by examining neutrino mixing as a function of angle of incidence
measured by the Super-Kamiokande Collaboration\cite{SKI}.
It is usually assumed that differing
macroscopic media affect the various traversing massless fermions in a manner expressing dispersive differences
quantified by the index $n_s(\epsilon$) for fermion type `s' in a manner analogous to photons traversing polarizable media.
In terms of the space-like distance traversed by a helicity specified particle $\Delta z={c \over n_s} \Delta t$, the corresponding light-like affine coordinate change in (\ref{NeutrinoMixingEqn}) is then expected to be
$\Delta \xi=\sqrt{2} c \Delta t=\sqrt{2} n_s \Delta z$.


\setcounter{equation}{0}
\section{Discussion and Conclusions}
\indent

A model describing a comprehensive group of transformations for causal and (non-perturbatively)
cluster decomposable physical systems has been demonstrated. 
The group extends Lorentz/Poincare' transformations to include additional
generators such that the fundamental representation states satisfy the Dirac equation. 
The general normalized spinors associated with causal fields
of group representation
labeled by $\Gamma$ (which is a half-integer or integer)  are Lorentz
transformed forms of the simplest spinors associated with the standard state reference frame. 

The spinors of the fundamental $\Gamma=\half$ representation
transform under a subgroup within GL(4).  The remaining 12 hermitian generators of GL(4)
can be chosen to consist of a subgroup
of 4 generators of SU(2)$\times$U(1) that leave the helicity states of massive fermions invariant,
with an additional set of 8 generators (including only 4 of the generators of SU(3))
that do not form a closed subgroup.  Alternatively, if
SU(3) is chosen as a subgroup of transformations,
the mass-SU(2) states have been shown to \emph{necessarily} mix within this subgroup. 
A justification for establishing (and relating)
either of these sets of subgroups as local gauge groups associated with \emph{charges}
has been presented.

Furthermore, since the SU(2)$\times$U(1) and SU(3) subgroups are necessarily related within the GL(4) group structure,
it is expected that relationships between \emph{quantized} charges associated with the invariance subgroups
of the local gauge fields can be established in terms of the topologically stable mapping of the local transformation
parameters $\alpha^r(x)$ onto compact space-time surfaces (via the circulation or flux of
gradients $\vec{\nabla}\alpha^r(\vec{x})$) analogous to what occurs with superfluids.
The quantization criteria should be calculable once all Lie structure functions of the subgroups
are developed from group multiplication properties, which has only partially been done\cite{JLxLorentz}.

An invariant SU(3) subgroup cannot be constructed for \emph{massless} fundamental fermions.  However, these
fermions \emph{can} be labeled with differing standard state labels (in contrast to photons), allowing coherent
mixing of differing states without assigning differing small masses, thus maintaining exact Lorentz invariance.  A
formulation providing measurable parameters associated with coherent propagation of massless
fundamental representation fermions through media has been presented.


\section*{Acknowledgments}
The author wishes to acknowledge the Visiting Faculty Program at Brookhaven National
Laboratory, and in particular Mickey Chiu for collaborating in exploring various
popular models of neutrino mixing.


\end{document}